\documentclass[bimj,fleqn]{w-art}
\usepackage{times}
\usepackage{w-thm}
\usepackage[authoryear]{natbib}
\theoremstyle{plain}

\theoremstyle{definition}

\usepackage[]{graphicx}
\chardef\bslash=`\\ 

\usepackage{color}

\begin{document}
\DOIsuffix{bimj.DOIsuffix}
\Volume{XX}
\Issue{YY}
\Year{2018}
\pagespan{1}{}

\title[A Bayesian perspective]{Contribution to the discussion of ``When should meta-analysis avoid making hidden normality assumptions?'': A Bayesian perspective}
\author[C.~R\"{o}ver]{Christian R\"{o}ver\footnote{Corresponding author: {\sf{e-mail: christian.roever@med.uni-goettingen.de}}, Phone: +49-551-395969, Fax: +49-551-394995}\inst{,1}} 
\address[\inst{1}]{Department of Medical Statistics, University Medical Center G\"{o}ttingen, Humboldtallee~32, 37073~G\"{o}ttingen, Germany}
\author[T.~Friede]{Tim Friede\inst{1}}

\Receiveddate{zzz} \Reviseddate{zzz} \Accepteddate{zzz}

\maketitle                   







  We congratulate Drs.\ Jackson and White on a very interesting paper,
  providing a comprehensive summary of explicit and implicit normality
  and independence assumptions commonly being made in random-effects
  meta-analysis that are easily overlooked and that may deserve more
  consideration \citep{JacksonWhite2018}.
  Some of the problems discussed, however, are less of an issue when
  analyses are done in a Bayesian framework. Normality assumptions are
  made explicit in the model's likelihood specification. Any departure
  from assumptions 1--4 (as listed in Tab.~3) then may cause problems,
  just as in the frequentist case.  Additional assumptions enter the
  analysis in the form of the prior specification, expressing the
  information on parameters before taking the data into consideration.
  These can usually be made reasonably vague, if desired (e.g., a
  uniform prior for the effect~$\mu$, and a weakly informative prior
  for the heterogeneity~$\tau$), and these may also be subjected to
  sensitivity analyses \citep{Roever2017}.  At the inference or
  prediction stage, computations are usually carried out
  \emph{exactly}, following Bayes' theorem, and, unlike in the
  frequentist framework, no approximations are necessary at this point
  \citep[Sec.~3]{SpiegelhalterEtAl}.
  Normal or other approximations \emph{could} of course also be used
  in a Bayesian analysis, e.g., when extrapolating around the posterior
  mode \citep[Sec.~13]{BDA3rd} or when when analysis is based on INLA
  \citep{RueMartinoChopin2009}. Such methods have been proposed for
  network meta-analysis \citep{SauterHeld2015,GuenhanEtAl2018},
  however, such cases are usually explicitly indicated.

\begin{figure}[htb]  
    \begin{center}
      \includegraphics[width=0.45\linewidth]{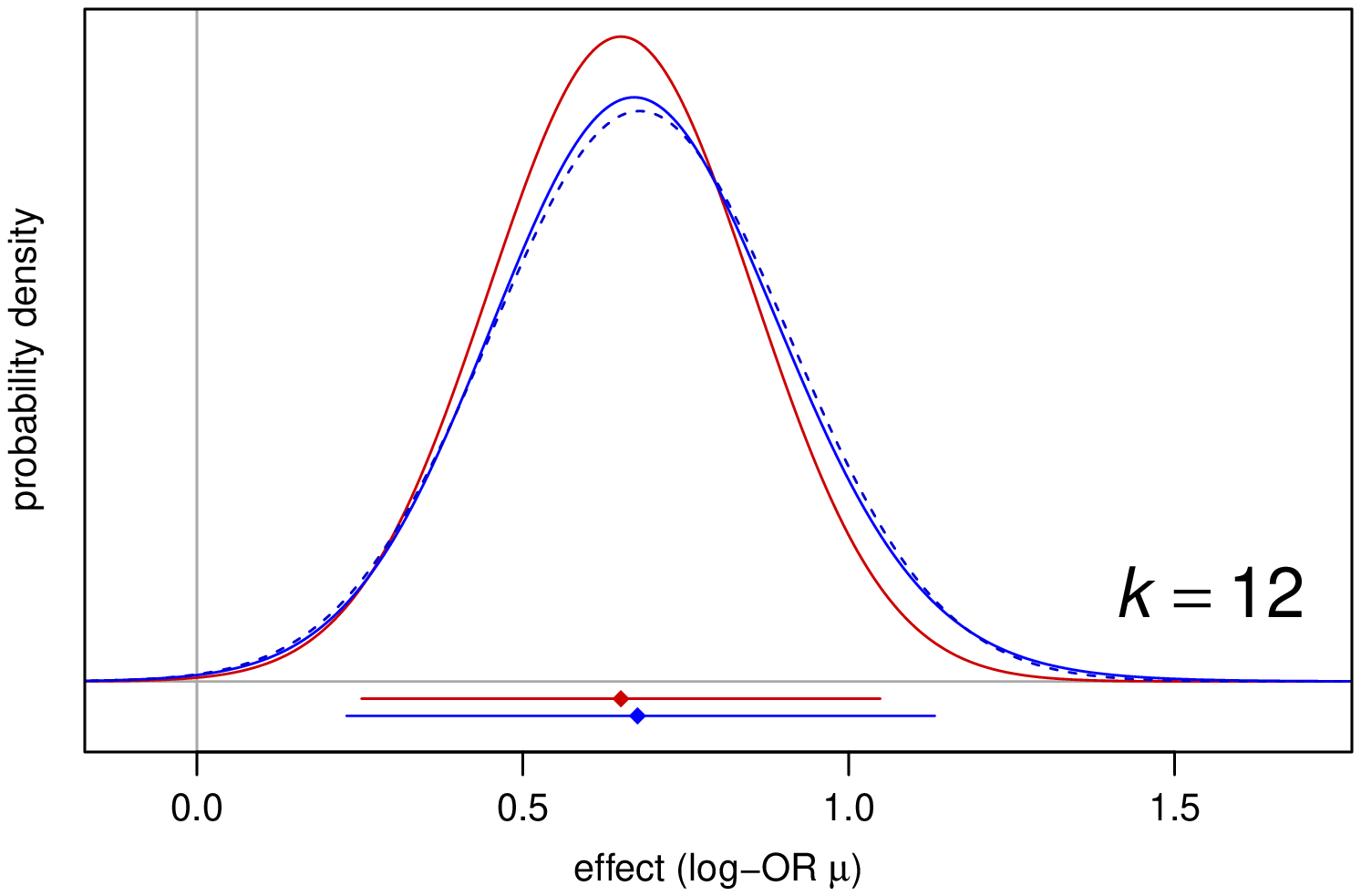}
      \hspace{0.05\linewidth}
      \includegraphics[width=0.45\linewidth]{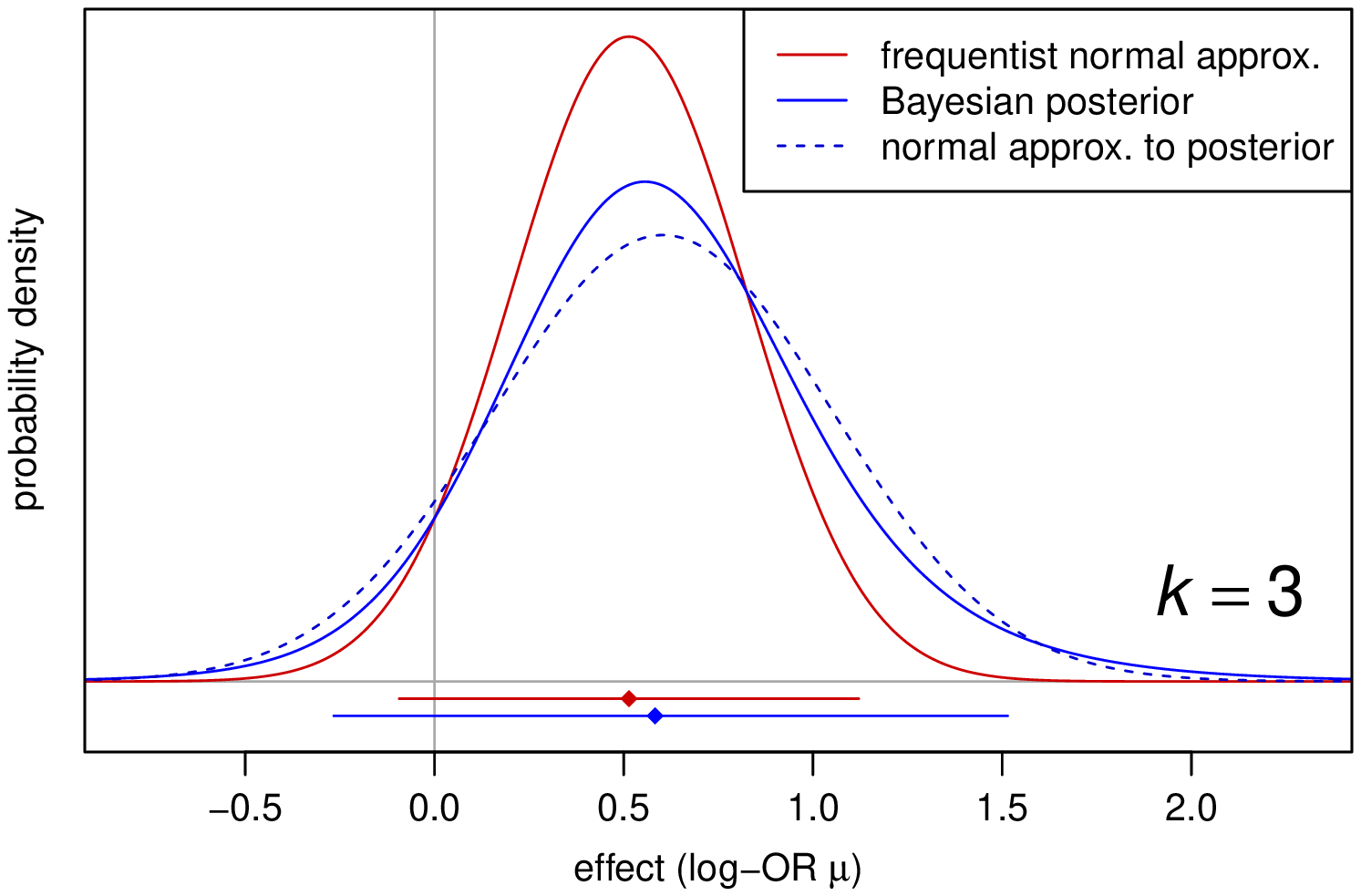} 
      \caption{\label{fig:densities}While the frequentist confidence
        interval is based on a normal approximation (red), the
        posterior (blue) is not normal. For comparison, a normal
        approximation to the posterior (with matching mean and
        variance) is also shown (dashed line).  Discrepancies are even
        more evident when the analysis is based on a smaller subset of
        only $k\!=\!3$~most recent studies (right panel).
        Corresponding estimates and 95\% confidence/credible intervals
        are shown at the bottom of the plot.}
    \end{center}
  \end{figure}

  To exemplify the issue, consider the smoking cessation example data
  (example one).  We utilize the \texttt{bayesmeta} \textsf{R}~package
  to derive the posterior distribution for the logarithmic odds ratio
  (log-OR), using an (improper) uniform prior for the effect~$\mu$ and
  a half-normal prior with scale~0.5 for the heterogeneity~$\tau$
  \citep{Roever2017}.  For the frequentist analyses, we use the
  \texttt{metafor} package with default settings
  \citep{Viechtbauer2010}.  Figure~\ref{fig:densities} illustrates the
  normal approximation utilized in the construction of the confidence
  interval along with the effect's posterior density. The (marginal)
  posterior distribution is not normal, but rather a \emph{normal
    mixture} \citep{Roever2017}, as becomes obvious when comparing to
  a normal approximation (based on matching mean and variance) that is
  also shown.

  Differences between (frequentist) normal approximation and posterior
  tend to be particularly large in case of substantial uncertainty in
  the heterogeneity parameter. While this happens especially in the
  common case of few studies
  (\citealp{FriedeRoeverWandelNeuenschwander2017a}; see also
  Fig.~\ref{fig:densities}), differences are still noticeable in the
  present data set consisting of twelve seemingly homogeneous OR
  estimates, where, due to an estimated heterogeneity of
  $\hat{\tau}\!=\!0$, effectively one ends up performing a
  common-effect analysis in the frequentist framework. An value of
  $\hat{\tau}\!=\!0$ is a ``most optimistic'' estimate here, in the
  sense that it will lead to the shortest possible confidence interval
  for a random-effects analysis.

  Although the data seem to ``look'' homogeneous, and despite the fact
  that we have a somewhat large sample size of $k\!=\!12$~studies, there
  is rather little information on the actual heterogeneity level in
  the data.  From the investigation by \citet{TurnerEtAl2015}, we know
  that for a general Cochrane review, empirically we may expect a
  heterogeneity level of roughly up to $\tau\!=\!1.16$ with 95\%
  probability. In the present example, the (Q-profile) 95\% confidence
  interval for $\tau$ is [0.00, 0.95], so we cannot say that the data
  provided a substantial constraint on the actual heterogeneity
  level. Although the $\tau$~estimate is zero and the data are
  compatible with a common-effect model, they are also consistent with
  even ``\emph{fairly high}'' heterogeneity
  \citep[Sec.~5.7.3]{SpiegelhalterEtAl}.  

  A Bayesian random-effects analysis is able to consider the range of
  plausible heterogeneity values via marginalization over the joint
  posterior distribution and with that leads to slightly more cautious
  results here. Due to the large standard errors, the relative
  increase in CI width is not dramatic in this case, but the Bayesian
  analysis yields an estimated log-OR and 95\% CI of 0.68 [0.23, 1.13]
  compared to 0.65 [0.25, 1.05] from the frequentist analysis.  While
  the half-normal heterogeneity prior with scale~$0.5$ focuses the
  analysis a~priori on up to \emph{fairly high} heterogeneity values
  (mostly $\tau\leq 1.0$), a sensitivity analysis using a scale
  of~$1.0$ leads to a similar estimate of 0.68 [0.22, 1.16].

  Since no \mbox{large-$k$-} (many-study-) asymptotics are invoked in
  a Bayesian analysis, the results are valid independent of the number
  of included studies. This is particularly useful in the case of a
  meta-analysis of few studies only, which is very common expecially
  in the context of medical applications: a majority of studies
  published in the \emph{Cochrane Library} are based on as few as
  $2$--$3$ studies
  \citep{DaveyEtAl2011,KontopantelisSpringateReeves2013}.  While
  small-$k$ adjustments for confidence intervals based on
  Student\mbox{-$t$} distributions are available, these only work
  really well in certain cases (equal standard errors) and tend to
  lead to unnecessarily wide intervals for few studies
  \citep{RoeverKnappFriede2015,FriedeRoeverWandelNeuenschwander2017a,FriedeRoeverWandelNeuenschwander2017b}.

  In a Bayesian model, it is generally relatively easy to accommodate
  model variations that avoid simplifying approximations \emph{at the
    modeling stage} (such as unknown standard errors, non-normal
  study-level likelihood, or non-normal heterogeneity), especially
  when applying MCMC methods \citep[e.g.,][]{Stevens2011}.
  In the frequentist case, usually dedicated GLMM software is required
  for this purpose. Incorporation e.g. of standard error uncertainty
  is less straightforward \citep{DominguezIslasRice2018}.  Quite
  often, more complex models may also lead to numerical difficulties
  especially in the common case of a small number of observations
  \citep[e.g.,][]{JacksonEtAl2018,SeideEtAl2018a}, so that application of certain
  more sophisticated models may simply not be possible in some
  instances.
  Even if more detailed likelihood specifications are used, the
  inference stage will commonly still rely on approximations (e.g.,
  via Wald-type intervals or likelihood ratio tests), which may in
  fact be even more questionable in more complex models.

  We would like to thank Drs.\ Jackson and White for an inspiring paper which highlighted some important issues in random-effects meta-analyses. We feel that some of the issues are less problematic in the Bayesian framework and recommend Bayesian random-effects meta-analyses for practical applications.

\vspace*{1pc}

\noindent {\bf{Conflict of Interest}}

\noindent {\it{The authors have declared no conflict of interest. 
           }}

%
%

{
  \bibliographystyle{bimj}
  \bibliography{/home/christian/literature/literature}
}
\end{document}